\documentclass[11pt]{article}
\usepackage{amsmath}
\usepackage{amssymb}
\usepackage{amsthm}
\usepackage{amsfonts}
\usepackage{comment}
\usepackage{color}
\usepackage{mathrsfs}
\usepackage{braket}
\usepackage{graphicx}
\usepackage[subrefformat=parens]{subcaption}
\usepackage{cite}
\usepackage{here}

\usepackage[stable]{footmisc}

\makeatletter

\@addtoreset{equation}{section}
\makeatother

\begin{document}


\title{\bf{Universality on thermodynamic relation with corrections in de Sitter black holes}}

\date{}
\maketitle
\begin{center}

{\large 
Junbeom Ko$^\clubsuit$\footnote{junbum9817@gmail.com}, Bogeun Gwak$^\clubsuit$\footnote{rasenis@dgu.ac.kr}
} \\
\vspace*{0.5cm}

{\it 
$^\clubsuit$Department of Physics, Dongguk University, Seoul 04620, Republic of Korea
}

\end{center}

\vspace*{1.0cm}

\begin{abstract}
{\noindent
We herein investigate the universal relation proposed by Goon and Penco in de Sitter black holes with electric charge or angular momentum. Our analysis focuses on the cosmological horizon, which only exists in de Sitter and Nariai spacetimes. Because the relation is given in a general case, the overall relationship may be valid. However, we elucidate the details of the relation, highlighting distinctions from those of (anti-)de Sitter black holes while affirming the validity of the relation. Furthermore, based on our analysis of Schwarzschild--de Sitter, Reissner--Nordstr\"om--de Sitter, and Kerr--de Sitter black holes, we demonstrate the universality of the thermodynamic relation in de Sitter black holes.
}
\end{abstract}

\newpage
\baselineskip=18pt
\setcounter{page}{2}
\pagestyle{plain}
\baselineskip=18pt
\pagestyle{plain}

\setcounter{footnote}{0}
	
\section{Introduction}
Black holes are one of the compact objects in general relativity, and they represent the final stage of the collapse of massive stars. They possess a spherical surface known as the event horizon, which has no outgoing geodesic. Upon entering the horizon, an ingoing geodesic moves toward the center of the black hole, recognized as the singularity \cite{Penrose:1964wq}. Consequently, no radiation can escape black holes in any manner. From a classical viewpoint, the mass of a black hole comprises reducible and irreducible components. The reducible energy can only be altered via interactions, whereas the irreducible mass does not decrease \cite{Christodoulou:1970wf, Christodoulou:1971pcn}. Hawking introduced the concept of black holes radiating energy, known as Hawking radiation. This insight enables black holes to be conceptualized as thermal objects with a temperature proportional to surface gravity \cite{Hawking:1974rv, Hawking:1975vcx}. These characteristics of black holes prompted the formulation of entropy by Bekenstein\cite{Bekenstein:1973ur, Bekenstein:1974ax}, which is related to the surface area of the black holes and is referred to as Hawking--Bekenstein entropy. Following the definition of the relevant variables, the laws of black hole thermodynamics were established in 1970s \cite{Bardeen:1973gs}. These laws encompass the zeroth, first, second, and third laws, grounded in concepts such as surface gravity over the event horizon of stationary black holes, fundamental form of the first law, and assertion of the non-decreasing area of black holes. Additionally, there was a mass formula based on the mass bound and constant black hole area of the Kerr--Newman black hole \cite{Smarr:1972kt}. These studies have transformed black holes into thermally radiating objects, with their behavior explained by thermodynamic variables and laws.

(Anti-)de Sitter ((A)dS) spacetime is a maximally symmetric Lorentzian manifold. The spacetime with a positive cosmological constant, $\Lambda$, is referred to as the de Sitter (dS) spacetime, whereas the spacetime with a negative constant is referred to as the AdS spacetime. In contrast to the Minkowski space, the submanifold of the dS spacetime is a hyperboloid in a five-dimensional flat spacetime. Coordinate singularities occur at $\xi=0, \xi=\pi, \theta=0,$ and $\theta=\pi$ when using the coordinate representation $(t,\xi,\theta,\phi)$. Furthermore, the spatial sections with a fixed $t$ take the form of spheres $S^3$ with a positive curvature. However, the AdS spacetime can be represented as a coordinate patch of half of the Einstein static universe. Gravity in the AdS spacetime is closely related to conformal field theory (CFT) on its boundary. This relation is  the AdS/CFT correspondence \cite{Maldacena:1997re, Witten:1998qj}, which asserts a relationship between $D$-dimensional AdS spacetime and a $(D-1)$-dimensional CFT defined on its boundary. In the context of AdS correspondence, efforts have been directed toward establishing a similar correspondence in the dS spacetime, known as the dS/CFT correspondence, linking the dS spacetime to the CFT of its boundary \cite{Strominger:2001pn, Strominger:2001gp}.

In the realm of quantum gravity, Vafa proposed the Weak Gravity Conjecture (WGC) as an explanatory framework for the charge-to-mass ratio \cite{Vafa:2005ui, Arkani-Hamed:2006emk}. Succinctly stated, the WGC posits that any gauge force must exert a greater influence than gravity, expressed as $\frac{Q}{M} \geq 1$, with saturation occurring in extremal black holes. This conjecture is rooted in quantum gravity principles and is particularly relevant in the absence of global symmetries. In the context of quantum gravity, as Hawking radiation does not entail the emission of charged particles, evaporating black holes release particles regardless of their global charges \cite{Banks:2010zn, Harlow:2022ich}. The WGC encompasses a group of studies that align with the ideas related to this conjecture. Various methods support the conjecture, including studies introducing correction terms to avoid naked singularities, as the absence of naked singularities implies the validity of the WGC inequality \cite{Cheung:2018cwt}.

Building on studies on the WGC, Goon and Penco investigated the universality of the thermodynamic relation between entropy and extremality under perturbation \cite{Goon:2019faz}. Perturbations in free energy yield a relationship between mass, temperature, and entropy with corrections. The leading-order expansion of perturbative parameters reveals an approximate relation that can be linked to higher-derivative corrections\cite{Wang:2022sbp}, generating a connection between shifts in entropy and the charge-to-mass ratio \cite{Cheung:2018cwt, Cheung:2019cwi, Kats:2006xp}. This relation ensures the validity of the WGC when the shift in the mass of extremal black holes is proportional to the shift of entropy with a negative constant \cite{Reall:2019sah} which is also studied in \cite{Ma:2023qqj}. Based on the relation proposed by Goon and Penco, considerable progress has been made, including analysis of the relation between various AdS spacetimes such as charged BTZ black holes and Kerr--AdS black holes from the WGC perspective \cite{Cano:2019ycn,Cremonini:2019wdk,Cano:2019oma,Sadeghi:2020xtc, Wei:2020bgk, Chen:2020rov, Chen:2020bvv, Sadeghi:2020ciy,Ma:2020xwi,McPeak:2021tvu,McInnes:2021frb,Etheredge:2022rfl,Sadeghi:2022xcr}.

In this work, we investigate the universal relation proposed by Goon and Penco in dS black holes \cite{Goon:2019faz}, where it has not been extensively studied in dS spacetimes. In particular, our focus is on another extremal condition where the outer horizon and cosmological horizon coincide, known as the Nariai spacetime. While the relation is proposed in a general case, we investigate the details of the equations that constitute the relation. The components of the relations are observed to differ from those of AdS black holes, but the overall relation remains valid. By analyzing different types of black holes, namely Schwarzschild--de Sitter (SdS), Reissner--Nordstr\"om--de Sitter (RNdS), Kerr--de Sitter (KdS), and Kerr--Newman--de Sitter (KNdS), the universality of the thermodynamic relation is confirmed in dS black holes. Furthermore, our results suggest that the WGC is still applicable in the dS spacetime.

The structure of this paper is as follows: In Section 2, we provide a review of thermodynamic variables and the Nariai limit of the three types of black holes. In Section 3, we examine \eqref{GP} in SdS black holes. Subsequently, we propose the application of the relation to RNdS black holes in Section 4 and KdS and KNdS black holes in Section 5. Finally, we summarize our findings in Section 6.

\section{Universal Relation and Nariai Limit}
\subsection{Goon and Penco Relation}
Goon and Penco built the relation between the derivative of mass and entropy \cite{Goon:2019faz},
    \begin{align}
        \frac{\partial M_{ext}(\Vec{\mathcal{Q}},\epsilon)}{\partial\epsilon}=\lim_{M \to M_{ext}(\Vec{\mathcal{Q},\epsilon})} -T\left(\frac{\partial S(M,\Vec{\mathcal{Q}},\epsilon)}{\partial\epsilon}\right)_{M,\Vec{\mathcal{Q}}},
    \label{GP}
    \end{align}
where $M$, $\Vec{\mathcal{Q}}$, and $\epsilon$ represent mass, additional quantities, and the perturbative parameter, respectively. While introducing perturbative corrections to the free energy $G(T,\Vec{\mu})$, including a proportional correction to the action $I$ of the system is viable. While contributions by boundary terms may arise, they are negligible. This correction is specifically applied to the cosmological term in the action, resulting in alterations to the horizon radius, mass, temperature, and entropy. Furthermore, the leading-order expansion of \eqref{GP} can be linked to the Weak Gravity Conjecture (WGC) by establishing proportions between the higher derivatives of mass and entropy.
    \begin{align}
        \Delta M_{ext}(\Vec{\mathcal{Q}})\approx -T_0(M,\Vec{\mathcal{Q}})\Delta S(M,\Vec{\mathcal{Q}})|_{M\approx M_{ext}^0(\Vec{\mathcal{Q}})}.
    \label{GPapp}
    \end{align}
This expansion is associated with higher-derivative corrections, giving rise to $\Delta S(M,\mathcal{Q})\sim\Delta z>0$ as demonstrated in \cite{Cheung:2018cwt}. Importantly, if our calculations align well with \eqref{GP}, it expands \eqref{GPapp}, demonstrating the validity of the WGC.

\subsection{SdS Black Hole}
In the case of dS spacetime, the metric includes additional terms related to the cosmological horizon, characterized by the cosmological constant $\Lambda$. The horizons, denoted as the inner, $r_i$, outer, $r_o$, and cosmological, $r_c$ horizons ($r_i<r_o<r_c$), can be derived. However, the Schwarzschild black hole, with only two horizons, possesses only $r_o$ and $r_c$. To obtain the extremal limit, the horizon of the black hole must be equal to both the inner and outer horizons, i.e., $r_h \equiv r_i=r_o$, accompanied by a temperature of zero. In contrast, the Nariai limit has a horizon identical to both the outer and cosmological horizons, i.e., $r_h=r_o=r_c$. For simplicity, we assume $G=c=1$. SdS black holes are spherically symmetric solutions of Einstein’s equations with zero electric charge and angular momentum. The metric is expressed as follows
    \begin{align}
        \begin{split} &
            ds^2=-f(r)dt^2+\frac{1}{f(r)}dr^2+r^2\left(d\theta^2+\sin^2\theta d\phi^2\right), \quad
            f(r)=1-\frac{2M}{r}-\frac{\Lambda r^2}{3}.
        \end{split}
    \label{eq:sch_ori}
    \end{align}
where M and $\Lambda$ represent the mass of the black hole and cosmological constant, respectively ($\Lambda=\frac{3}{\ell^2}$. The entropy and Hawking temperature at the event horizon $r_h$ are given by
    \begin{align}
        \begin{split}
            S=\frac{A_h}{4}=\pi r_h^2,\quad
            T=\frac{\kappa}{2\pi}=\frac{1}{4\pi}\left(\frac{\partial f}{\partial r}\right)_{r=r_h},
        \end{split}
    \label{eq:sch_oT}
    \end{align}
where $A_h$ is the surface area of the black hole. Utilizing these thermodynamic variables, including the mass of the black hole, we can formulate the first law of thermodynamics as
    \begin{align}
        dM=TdS.
    \end{align}
If we get $f(r)=f'(r)=0$, we can derive the Nariai radius and mass as
    \begin{gather}
        r_N=\frac{l}{\sqrt{3}}, \\ 
        M_N=\frac{l}{3\sqrt{3}}.
    \label{eq:sch_NM}
    \end{gather}

\subsection{RNdS Black Hole}

The RNdS black hole is a static solution of the Einstein--Maxwell equations, introducing an additional term for electric charge over that present in the SdS case:
    \begin{align}
        \begin{split} 
            ds^2&=-f(r)dt^2+\frac{1}{f(r)}dr^2+r^2\left(d\theta^2+\sin^2\theta d\phi^2\right), \\ 
            f(r)&=1-\frac{2M}{r}+\frac{Q^2}{r^2}-\frac{1}{3}\Lambda r^2,\quad
            A_\mu dx^\mu=\frac{iQ}{r}dt,
        \end{split}
    \label{eq:rn_ori}
    \end{align}
where $Q$ represents the electric charge. The RNdS black holes share the same form of entropy and temperature as Schwarzschild black holes. Evaluating all variables in RNdS black holes enables us to establish the first law of thermodynamics:
    \begin{align}
        dM=TdS+\Phi dQ.
    \end{align}
We then compute the radius and mass as those for Schwarzschild black holes,
    \begin{gather}
        r_N=\frac{l}{\sqrt{6}}\sqrt{1+\sqrt{1-\frac{12Q^2}{l^2}}}, \\ 
        M_N=\frac{12Q^2+l^2\left(1+\sqrt{1-\frac{12Q^2}{l^2}}\right)}{3\sqrt{6}l\sqrt{1+\sqrt{1-\frac{12Q^2}{l^2}}}}.
    \label{eq:rn_NM}
    \end{gather}

\subsection{KdS Black Hole}

The metric for KdS Black holes represents one of the stationary solutions of the Einstein equations. The KdS metric expressed in Boyer--Lindquist coordinates is as follows as 
    \begin{align}
        \begin{split} &
            ds^2=-\frac{\Delta_r}{\rho^2}\left(dt-\frac{a\sin^2\theta}{\Xi}d\phi^2\right)+\frac{\rho^2}{\Delta_r}dr^2+\frac{\rho^2}{\Delta_\theta}d\theta^2+\frac{\Delta_\theta \sin^2\theta}{\rho^2}\left(adt-\frac{r^2+a^2}{\Xi}d\phi\right)^2, \\ &
        \Delta_r=\left(r^2+a^2\right)\left(1-\frac{1}{3}\Lambda r^2\right)-2mr, \quad 
        \Delta_\theta=1+\frac{1}{3}a^2 \cos^2\theta, \\ &
        \rho^2=r^2+a^2\cos^2\theta, \quad 
        \Xi=1+\frac{1}{3}\Lambda a^2,
        \end{split}
    \label{eq:k_ori1}
    \end{align}
where $m$ and $a$ represent the mass and spin parameter, respectively. Before deriving thermodynamic variables, we must transform \eqref{eq:k_ori1} into another form. This transformation is necessary because it has a non-zero angular velocity at the boundary $r\gg1$, rendering the first law invalid. To address this issue, we must perform a transformation as outlined in \cite{Hawking:1998kw, Gwak:2018akg, Ponglertsakul:2020ufm}.
    \begin{align}
        t\rightarrow T,\quad \phi\rightarrow\Phi+\frac{1}{3}a\Lambda T.
    \end{align}
We can then change \eqref{eq:k_ori1} into\cite{Gwak:2021tcl},
    \begin{align}
        \begin{split} 
        ds^2=&-\frac{\Delta_r}{\rho^2\Xi^2}\left(\Delta_\theta dT-a\sin^2\theta d\Phi\right)^2+\frac{\rho^2}{\Delta_r}dr^2+\frac{\rho^2}{\Delta_\theta}d\theta^2 \\ 
        &+\frac{\Delta_\theta\sin^2\theta}{\rho^2\Xi^2}\left(a\left(1-\frac{1}{3}\Lambda r^2\right)dT-\left(r^2+a^2\right)d\Phi\right)^2.
        \end{split}
    \end{align}
KdS black holes exhibit a different form of entropy and temperature \cite{Gregory:2021ozs} compared to the other two black holes,
    \begin{align}
        S_h=\frac{\pi\left(r_h^2+a^2\right)}{\Xi},\quad  T_h=\left|\frac{\partial_r\Delta_r|_{r=r_h}}{4\pi\left(r_h^2+a^2\right)}\right|.
    \label{eq:k_S}
    \end{align}
Additionally, other variables include the actual mass, angular momentum, and angular velocity \cite{Caldarelli:1999xj}.
    \begin{align}
        M=\frac{m}{\Xi^2},\quad J=\frac{am}{\Xi^2},\quad \Omega_h=\frac{a\left(1-\frac{1}{3}\Lambda r_h^2\right)}{r_h^2+a^2}.
    \label{eq:k_va}
    \end{align}
From the thermodynamic variables derived previously, we can formulate the first law of thermodynamics for KdS black holes as
    \begin{align}
        dM=T_hdS_h+\Omega_h dJ.
    \label{eq:fl}
    \end{align}
Next, we calculate the Nariai radius and set the mass variation $\Delta_r$ in \eqref{eq:k_ori1} to zero.
    \begin{gather}
        r_N=\frac{\sqrt{l^2-a^2}}{\sqrt{6}}\sqrt{1+\sqrt{1-\frac{12a^2 l^2}{\left(l^2-a^2\right)^2}}}, \\
        M_N=\frac{30a^2l^2+\left(l^2-a^2\right)^2\left(1+\sqrt{1-\frac{12a^2l^2}{\left(l^2-a^2\right)^2}}\right)}{6\sqrt{6}l^2\sqrt{l^2-a^2}\sqrt{1+\sqrt{1-\frac{12a^2l^2}{\left(l^2-a^2\right)^2}}}}.
    \label{eq:k_NM}
    \end{gather}
The mass can then be derived by substituting \eqref{eq:k_ori1} into the mass function of the KdS black hole.

\section{Universal Relation on SdS Black Hole}
\noindent
We will evaluate \eqref{GP} for the three black holes, focusing on the Nariai limit, as opposed to the methods followed in the investigations by Goon and Penco \cite{Goon:2019faz}. We begin with the SdS case. First, we examine how the action changes:
	\begin{align}
		I=\frac{1}{16\pi}\int dx^4 \sqrt{-g} \left(R-2\Lambda\right),
	\end{align}
where $R$ and $\Lambda$ are the Ricci scalar and cosmological constant. According to \cite{Goon:2019faz}, the cosmological constant becomes the perturbation parameter, yielding
    \begin{align}
        I=\frac{1}{16\pi}\int dx^4 \sqrt{-g} \left(R-2\left(1+\epsilon\right)\Lambda\right),
    \end{align}
where $\epsilon$ is a very small constant. The correction shifts the metric function in \eqref{eq:sch_ori},
    \begin{align}
        f(r)=1-\frac{2M}{r}-\frac{1+\epsilon}{3}\Lambda r^2=1-\frac{2M}{r}-\frac{(1+\epsilon)r^2}{l^2},
    \label{eq:sch_mf}
    \end{align}
where $\Omega^2_2=d\theta^2+\sin^2\theta d\phi$. We obtain the shifted mass set $f(r)=0$ in \eqref{eq:sch_mf},
    \begin{gather}
        M=\frac{\sqrt{S}}{2\sqrt{\pi}}-\frac{1+\epsilon}{2l^2}\frac{S^{\frac{3}{2}}}{\pi^{\frac{3}{2}}}.
    \label{eq:sch_M}
    \end{gather}
We need function of $\epsilon$ to build \eqref{GP}. This can be derived by transforming \eqref{eq:sch_M} as
    \begin{align}
        \epsilon=\frac{2l^2\pi^{\frac{3}{2}}}{S^{\frac{3}{2}}}\left(-M+\frac{\sqrt{S}}{2\sqrt{\pi}}\right){-1}=\frac{-2l^2\pi^{\frac{3}{2}}M}{S^\frac{3}{2}}+\frac{\pi l^2}{S}-1.
    \label{eq:sch_ep}
    \end{align}
Next, we obtain a partial derivative to \eqref{eq:sch_ep} with a fixed mass,
    \begin{align}
        \left(\frac{\partial \epsilon}{\partial S}\right)_M=\frac{\pi l^2-3\left(1+\epsilon\right)S}{2S^2}.
    \label{eq:sch_paep}
    \end{align}
We can compute temperature from \eqref{eq:sch_mf} using the Hawking temperature equation \eqref{eq:sch_oT},
    \begin{align}
        T=\frac{1}{4\pi}\frac{\pi l^2-3\left(1+\epsilon\right)S}{\sqrt{\pi S}}.
    \label{eq:sch_T}
    \end{align}
Combining \eqref{eq:sch_paep} and \eqref{eq:sch_T} to get left-hand side of \eqref{GP}, we obtain 
    \begin{align}
        -T\frac{\partial S}{\partial\epsilon}=-\frac{S^\frac{3}{2}}{2\pi^\frac{3}{2} l^2}.
    \label{eq:sch_rh}
    \end{align}
We then calculate the Nariai extremal condition, referred to as $S_N$, from $T=0$ before completing the \eqref{GP},
    \begin{align}
        S_N=\frac{\pi l^2}{3\left(1+\epsilon\right)}.
    \label{eq:sch_N}
    \end{align}
We begin with using \eqref{eq:sch_N} at the Nariai limit \eqref{eq:sch_rh} to obtain
    \begin{align}
        \lim_{M \to M_N}\left(-T\frac{\partial S}{\partial \epsilon}\right)_M=-\frac{l}{6\sqrt{3}\left(1+\epsilon \right)^\frac{3}{2}},
    \label{eq:sch_rhcom}
    \end{align}
and then substitute \eqref{eq:sch_N} into \eqref{eq:sch_M},
    \begin{align}
        M_N=\frac{l}{3\sqrt{3\left(1+\epsilon \right)}}.
    \label{eq:sch_Mfinal}
    \end{align}
After differentiation, we obtain the form:
    \begin{align}
        \left(\frac{\partial M_N}{\partial \epsilon}\right)_T=-\frac{l}{6\sqrt{3}\left(1+\epsilon\right)^\frac{3}{2}}.
    \label{eq:sch_lf}
    \end{align}
Lastly, by comparing \eqref{eq:sch_rhcom} and \eqref{eq:sch_lf}, we observe that the form of \eqref{GP} is also verified in the dS case, similar to \eqref{GP}. However, this differs from the result in \cite{Goon:2019faz}. Our exact solutions in \eqref{eq:sch_rhcom} and \eqref{eq:sch_lf} have a minus sign, which arises due to the presence of a positive cosmological constant, a characteristic feature of dS spacetimes.

\section{Universal Relation on RNdS Black Hole}

In this section, we will examine the RNdS metric. RNdS black holes feature an additional term with the electromagnetic tensor in the action, making them a solution to the Einstein--Maxwell field equation with a cosmological constant. The metric undergoes perturbation to the cosmological constant term in the action, similar to the SdS black hole discussed in Section 3,
    \begin{gather}
        I=\frac{1}{16\pi}\int dx^4 \sqrt{-g} \left(R-2\left(1+\epsilon \right)\Lambda-F_{\mu\nu}F^{\mu\nu}\right).
    \end{gather}
where $F_{\mu\nu}=\partial_\mu A_\nu-\partial_\nu A_\mu$ pertains to the electromagnetic tensor. Similar to the case of SdS black holes, the perturbation parameter on the cosmological constant modifies \eqref{eq:rn_ori} as
    \begin{align}
        f(r)=1-\frac{2M}{r}+\frac{Q^2}{r^2}-\frac{1}{3}\left(1+\epsilon\right)\Lambda r^2.
    \label{eq:rn_fm}
    \end{align}
We can obtain mass as an equation of $\epsilon$, $S$, and $Q$ from the metric function in \eqref{eq:rn_fm},
    \begin{align}
        M=\frac{\sqrt{S}}{2\sqrt{\pi}}+\frac{Q^2\sqrt{\pi}}{2\sqrt{S}}-\frac{1+\epsilon}{2l^2}\frac{S^\frac{3}{2}}{\pi^\frac{3}{2}}.
    \label{eq:rn_M}
    \end{align}
Now, we transform \eqref{eq:rn_M} to $\epsilon$ and differentiate it with entropy $S$ with a fixed mass and electric charge,
    \begin{gather}
        \epsilon=\frac{-2l^2 \pi^\frac{3}{2} M}{S^\frac{3}{2}}+\frac{l^2 \pi}{S}+\frac{Q^2 l^2 \pi^2}{S^2}-1, \\ 
        \left(\frac{\partial \epsilon}{\partial S}\right)_{M,Q}=\frac{\pi l^2 S-3\left(1+\epsilon\right)S^2-Q^2\pi^2 l^2}{2S^3}.
    \label{eq:rn_paep}
    \end{gather}
As we did for the SdS metric, we can calculate the Hawking temperature $T$ from \eqref{eq:rn_fm},
    \begin{align}
        T=\frac{1}{4\pi}\frac{\pi l^2 S-3\left(1+\epsilon \right)S^2-Q^2\pi^2 l^2}{l^2\sqrt{\pi}S^\frac{3}{2}}.
    \label{eq:rn_T}
    \end{align}
Combining \eqref{eq:rn_paep} and \eqref{eq:rn_T}, we obtain
    \begin{align}
        -T\frac{\partial S}{\partial \epsilon}=-\frac{S^\frac{3}{2}}{2\pi^\frac{3}{2} l^2}.
    \label{eq:rn_rh}
    \end{align}
Setting $T=0$ from \eqref{eq:rn_T} aids in identifying the entropy of Nariai extremal case. Because the equation has two solutions, extremal and Nariai, we select the Nariai case.
    \begin{align}
        S_N=\frac{\pi l^2}{6\left(1+\epsilon \right)}\left(1+\sqrt{1-\frac{12\left(1+\epsilon \right)Q^2}{l^2}}\right).
    \label{eq:rn_N}
    \end{align}
Substituting \eqref{eq:rn_N} into \eqref{eq:rn_rh}, we obtain
    \begin{align}
        \lim_{M \to M_N}\left(-T\frac{\partial S}{\partial \epsilon}\right)_{M, Q}=-\frac{l}{12\sqrt{6\left(1+\epsilon \right)^\frac{3}{2}}}\left(1+\sqrt{1-\frac{12\left(1+\epsilon \right)Q^2}{l^2}}\right)^\frac{3}{2}.
    \label{eq:rn_rhcom}
    \end{align}
Subsequently, we determine the left side of \eqref{GP} by inserting \eqref{eq:rn_N} into \eqref{eq:rn_M},
    \begin{align}
        M_N=\frac{12\left(1+\epsilon \right)Q^2+l^2\left(1+\sqrt{1-\frac{12\left(1+\epsilon\right)Q^2}{l^2}}\right)}{3\sqrt{6\left(1+\epsilon \right)}l\sqrt{1+\sqrt{1-\frac{12\left(1+\epsilon \right)Q^2}{l^2}}}}.
    \label{eq:rn_Mfinal}
    \end{align}
For the last step, we differentiate \eqref{eq:rn_Mfinal} with $\epsilon$,
    \begin{align}
        \left(\frac{\partial M_N}{\partial \epsilon}\right)_{T, Q}=-\frac{l}{12\sqrt{6}\left(1+\epsilon \right)^\frac{3}{2}} \left(1+\sqrt{1-\frac{12\left(1+\epsilon \right)Q^2}{l^2}}\right)^\frac{3}{2}.
    \label{eq:rn_lf}
    \end{align}
Given that \eqref{eq:rn_rhcom} and \eqref{eq:rn_lf} share the same form, \eqref{GP} can also be applied to the case of RNdS black holes. The tendency to attach a minus sign is also observed, mirroring the behavior of SdS black holes.

We examined the relation within the Nariai limit. However, for direct applicability to the WGC, applying our calculations within the extremal limit is preferable, because the original WGC concept is rooted in the extremal limit of AdS black holes. Therefore, we derive the extremal limit of RNdS black holes by obtaining another solution for $T=0$ from \eqref{eq:rn_T},
    \begin{align}
        S_E=\frac{\pi l^2}{6\left(1+\epsilon \right)}\left(1-\sqrt{1-\frac{12\left(1+\epsilon \right)Q^2}{l^2}}\right).
    \label{eq:rn_ext}
    \end{align}
Before establishing the relation in equation \eqref{GP}, we must derive equations akin to \eqref{eq:rn_rhcom} and \eqref{eq:rn_lf}. By substituting \eqref{eq:rn_ext} into \eqref{eq:rn_rh}, we obtain
    \begin{align}
        \lim_{M \to M_E}\left(-T\frac{\partial S}{\partial \epsilon}\right)_{M, Q}=-\frac{l}{12\sqrt{6\left(1+\epsilon \right)^\frac{3}{2}}}\left(1-\sqrt{1-\frac{12\left(1+\epsilon \right)Q^2}{l^2}}\right)^\frac{3}{2}.
    \label{eq:rn_rhext}
    \end{align}
If we subsequently substitute \eqref{eq:rn_ext} into \eqref{eq:rn_M}, the mass can be expressed as
    \begin{align}
        M_E=\frac{12\left(1+\epsilon \right)Q^2+l^2\left(1-\sqrt{1-\frac{12\left(1+\epsilon\right)Q^2}{l^2}}\right)}{3\sqrt{6\left(1+\epsilon \right)}l\sqrt{1-\sqrt{1-\frac{12\left(1+\epsilon \right)Q^2}{l^2}}}}.
    \label{eq:rn_mext}
    \end{align}
Upon taking the partial derivative of \eqref{eq:rn_mext}, we obtain an equation similar to \eqref{eq:rn_lf},
    \begin{align}
        \left(\frac{\partial M_E}{\partial \epsilon}\right)_{T, Q}=-\frac{l}{12\sqrt{6}\left(1+\epsilon \right)^\frac{3}{2}} \left(1-\sqrt{1-\frac{12\left(1+\epsilon \right)Q^2}{l^2}}\right)^\frac{3}{2}.
    \label{eq:rn_lfext}
    \end{align}
Finally, we observe that \eqref{GP} remains valid in the extremal limit of RNdS black holes. However, both \eqref{eq:rn_rhext} and \eqref{eq:rn_lfext} feature a negative sign in their exact solutions, akin to the Nariai limit case. Thus, RNdS black holes in the extremal limit also adhere to the trend observed in RNdS black holes within the Nariai limit.

\section{Universal Relation on KdS Black Hole}

\subsection{Smarr Formula on KdS Black Hole}

The method for computing the mass or temperature of black holes should be clarified. The straightforward approach involves using the metric function, where $f(r)=0$. However, for the KdS black hole, calculating these quantities directly is not as straightforward owing to its complication. Therefore, we will resort to the Smarr formula. The original form can be found the Kerr--Newman black hole \cite{Smarr:1972kt},
    \begin{align}
        M=\left(\frac{A}{16\pi}+\frac{4\pi L^2}{A}+\frac{Q^2}{2}+\frac{\pi Q^4}{A}\right)^\frac{1}{2}.
    \end{align}
where $A$, $L$, and $Q$ represent the surface area, angular momentum, and electric charge, respectively. This form is generalized into AdS and dS black holes with a cosmological constant \cite{Caldarelli:1999xj,Sekiwa:2006qj}. Consequently, the generalized Smarr formula of the KdS black hole which we will use is given as
    \begin{align}
        M^2=\left(\frac{\pi}{S}-\frac{1}{l^2}\right)J^2+\frac{S}{4\pi}\left(1-\frac{S}{\pi l^2}\right)^2.
    \label{eq:Smarr}
    \end{align}

\subsection{Testing Universal Relation on KdS Black Hole}

Finally, we will verify the relation for KdS black holes. The KdS solution comprises axially symmetric black holes with an axis of rotation. The action is identical to the SdS action, leading to a process similar to the SdS case. $\Delta_r$ in \eqref{eq:k_ori1} then becomes:
    \begin{align}
        \Delta_r=\left(r^2+a^2\right)\left(1-\frac{1}{3}\left(1+\epsilon\right)\Lambda r^2\right)-2mr.
    \label{eq:k_mf}
    \end{align}
When $\epsilon=0$, \eqref{eq:k_mf} will be recovered to an uncorrected solution. With perturbation parameter $\epsilon$, we can derive mass from generalized Smarr formula in KdS black holes, \eqref{eq:Smarr},
    \begin{align}
        M=\sqrt{\frac{\pi l^2-\left(1+\epsilon\right)S}{4\pi^3 l^4 S}\left(4\pi^3l^2J^2+S^2\left(\pi l^2-(1+\epsilon)S\right)\right)},
    \label{eq:k_M}
    \end{align}
which can be reproduced as $M$ from $\Delta_r=0$ with $S=\frac{\pi (r^2+a^2)}{\Xi}$, $M=\frac{m}{\Xi^2}$, and $J=\frac{am}{\Xi^2}$. We then set \eqref{eq:k_M} as a function of $\epsilon$ to make the relation,
    \begin{align}
        \epsilon=\frac{\pi l^2\left(2\pi^2J^2+S^2-2\sqrt{\pi^4 J^4+\pi M^2 S^3}\right)}{S^3}-1.
    \label{eq:k_ep}
    \end{align}
Taking the partial derivative to \eqref{eq:k_ep}, we get
    \begin{align}
        \left(\frac{\partial \epsilon}{\partial S}\right)_{M,J}=\frac{-4\pi^4J^2l^4+S^2\left(\pi l^2-(1+\epsilon)S\right)\left(\pi l^2-3(1+\epsilon)S\right)}{4\pi^3J^2l^2S^2+2S^4\left(\pi l^2-(1+\epsilon)S\right)}.
    \label{eq:k_paep}
    \end{align}
In contrast to the other two sections, it is easier to calculate using \eqref{eq:fl} as $T=\left(\frac{\partial M}{\partial S}\right)_{J}$ \cite{Sekiwa:2006qj} and \eqref{eq:Smarr},
    \begin{align}
        T=\frac{-4\pi^4J^2l^4+S^2\left(\pi l^2-(1+\epsilon)S\right)\left(\pi l^2-3(1+\epsilon)S\right)}{4\pi^\frac{3}{2}l^2\sqrt{S^3\left(\pi l^2-(1+\epsilon)S\right)\left(4\pi^3J^2l^2+S^2(\pi l^2-(1+\epsilon)S)\right)}},
    \label{eq:k_T}
    \end{align}
which can be reproduced to the result of $T=\left|\frac{\partial_r \Delta_r}{4\pi \left(r^2+a^2\right)}\right|$ \cite{Gregory:2021ozs}. We can then build the right side of \eqref{GP} by combining \eqref{eq:k_paep} and \eqref{eq:k_T},
    \begin{align}
        -T\left(\frac{\partial S}{\partial\epsilon}\right)_{M, J}=-\frac{4\pi^3J^2l^2S^2+2S^4\left(\pi l^2-(1+\epsilon)S\right)}{4\pi^\frac{3}{2}l^2\sqrt{S^3\left(\pi l^2-(1+\epsilon)S\right)\left(4\pi^3J^2l^2+S^2(\pi l^2-(1+\epsilon)S)\right)}}.
    \label{eq:k_rh}
    \end{align}
Subsequently, we compute the Nariai extremal condition for KdS black holes, which can be calculated by setting $T=0$ because extremal black holes have zero temperature,
    \begin{align}
        S^2\left(\pi l^2-(1+\epsilon)\right)\left(\pi l^2-3(1+\epsilon)S\right)-4\pi^4J^2l^4=0.
    \label{eq:k_ec}
    \end{align}
If we solve \eqref{eq:k_ec}, the four roots will be obtained as
    \begin{align*}
        \begin{split} &
            S_{1,2}=\frac{\pi l^2}{3(1+\epsilon)}+\frac{1}{2}\sqrt{\frac{2\pi^2l^4}{9(1+\epsilon)^2}+P}\pm\frac{1}{2}\sqrt{\frac{4\pi^2l^4}{9(1+\epsilon)^2}-P+\frac{4\pi^3l^6}{27(1+\epsilon)^3\sqrt{\frac{2\pi^2l^4}{9(1+\epsilon)^2}+P}}}, \\ &
            S_{3,4}=\frac{\pi l^2}{3(1+\epsilon)}-\frac{1}{2}\sqrt{\frac{2\pi^2l^4}{9(1+\epsilon)^2}+P}\pm\frac{1}{2}\sqrt{\frac{4\pi^2l^4}{9(1+\epsilon)^2}-P-\frac{4\pi^3l^6}{27(1+\epsilon)^3\sqrt{\frac{2\pi^2l^4}{9(1+\epsilon)^2}+P}}},
        \end{split}
    \end{align*}
where
    \begin{align*}
        \begin{split} &
            P=\frac{\pi^2\left(l^8-144J^2l^4(1+\epsilon)^2\right)}{9(1+\epsilon)^2 Q^\frac{1}{3}}+\frac{\pi^2Q^\frac{1}{3}}{9(1+\epsilon)^2}, \\ &
            Q=l^{12}-432J^2l^8(1+\epsilon)^2+12\sqrt{3}\sqrt{-J^2l^{20}(1+\epsilon)^2+288J^4l^{16}(1+\epsilon)^4+6912J^6l^{12}(1+\epsilon)^6}.
        \end{split}
    \end{align*}
Moving forward, a detailed examination of the four roots obtained is necessary. If we consider the limit $J\rightarrow 0$, $S_1$, $S_3$, and $S_4$ do not revert to the SdS Nariai extremal condition. Additionally, the third term of $S_3$ and $S_4$ becomes imaginary as $J$ increases. Consequently, $S_2$ stands out as the only root suitable for use as $S_N$.
    \begin{align}
        S_N=S_2=\frac{\pi l^2}{3\epsilon}+\frac{1}{2}\sqrt{\frac{2\pi^2l^4}{9\epsilon^2}+P}-\frac{1}{2}\sqrt{\frac{4\pi^2l^4}{9\epsilon^2}-P+\frac{4\pi^3l^6}{27\epsilon^3\sqrt{\frac{2\pi^2l^4}{9\epsilon^2}+P}}}.
    \label{eq:k_N}
    \end{align}
Substituting \eqref{eq:k_N} into \eqref{eq:k_rh}, we then build the right hand side of the relation,
    \begin{align}
        \lim_{M \to M_N}\left(-T\frac{\partial S}{\partial\epsilon}\right)_M=-\frac{4\pi^3J^2l^2S_2^2+2S_2^4\left(\pi l^2-\epsilon S_2\right)}{4\pi^\frac{3}{2}l^2\sqrt{S_2^3\left(\pi l^2-(1+\epsilon)S_2\right)\left(4\pi^3J^2l^2+S_2^2(\pi l^2-(1+\epsilon)S_2)\right)}}.
    \label{eq:k_rhcom}
    \end{align}
We substitute \eqref{eq:k_N} into \eqref{eq:k_M}, 
    \begin{align}
        M_N=\sqrt{\frac{\pi l^2-\left(1+\epsilon\right)S_2}{4\pi^3 l^4 S_2}\left(4\pi^3l^2J^2+S_2^2\left(\pi l^2-(1+\epsilon)S_2\right)\right)},
    \label{eq:k_Mfinal}
    \end{align}
and take partial derivative with $\epsilon$,
    \begin{align}
        \left(\frac{\partial M_N}{\partial\epsilon}\right)_{T,J}=-\frac{4\pi^3J^2l^2S_2^2+2S_2^4\left(\pi l^2-\epsilon S_2\right)}{4\pi^\frac{3}{2}l^2\sqrt{S_2^3\left(\pi l^2-(1+\epsilon)S_2\right)\left(4\pi^3J^2l^2+S_2^2(\pi l^2-(1+\epsilon)S_2)\right)}}.
    \label{eq:k_lf}
    \end{align}
Note that \eqref{eq:k_rhcom} and \eqref{eq:k_lf} are identical. This implies that the GP relation can also be identified in KdS black holes with the Nariai limit. Moreover, a closer look reveals that the exact results in \eqref{eq:k_rhcom} and \eqref{eq:k_lf} feature a minus sign. This observation indicates that the trend observed in the properties of dS spacetime remains valid, thereby aligning with the findings in the other two types of black holes.

\subsection{Testing Universal Relation on KNdS Black Hole}

To apply our calculation to black holes with angular momentum and electric charge, we adjust \eqref{eq:k_mf} to represent the KNdS case, as follows
    \begin{align}
        \Delta_r=\left(r^2+a^2\right)\left(1-\frac{1}{3}\left(1+\epsilon\right)\Lambda r^2\right)-2mr+Q^2.
    \end{align}
Furthermore, we modify the Smarr formula, \eqref{eq:Smarr}, for KNdS black holes,
    \begin{align}
        M^2=\left(\frac{\pi}{S}-\frac{1}{l^2}\right)J^2+\frac{S}{4\pi}\left(\frac{\pi Q^2}{S}+1-\frac{S}{\pi l^2}\right)^2,
    \label{eq:kn_smarr}
    \end{align}
where Q is the electric charge. We can derive the perturbed mass from \eqref{eq:kn_smarr},
    \begin{align}
        M=\sqrt{\frac{1}{4\pi^3 l^4 S}\left(4\pi^3l^2J^2(\pi l^2-(1+\epsilon)S\right)+\left(\pi^2 Q^2 l^2+\pi l^2 S-(1+\epsilon)S^2\right)^2}.
    \label{eq:kn_m}
    \end{align}
By setting a function of $\epsilon$ from \eqref{eq:kn_m},
    \begin{align}
        \epsilon=\frac{\pi l^2\left(2\pi^2 J^2+\pi Q^2 S+S^2-2\sqrt{\pi^4 J^4+\pi^3 Q^2 J^2 S+\pi M^2 S^3}\right)}{S^3}-1.
    \label{eq:kn_epsilon}
    \end{align}
Then, taking the partial derivative of \eqref{eq:kn_epsilon} with respect to S,
    \begin{align}
        \left(\frac{\partial \epsilon}{\partial S}\right)_{M,J,Q}=\frac{-4\pi^4 J^2 l^4-\pi^2 Q^2 l^2\left(\pi^2 Q^2 l^2+2S^2\right)+S^2\left(\pi l^2-(1+\epsilon)S\right)\left(\pi l^2-3(1+\epsilon)S\right)}{4\pi^3 J^2 l^2 S^2+2\pi l^2 S^3(\pi Q^2+S)-2(1+\epsilon)S^5}.
    \label{eq:kn_paep}
    \end{align}
We can easily calculate temperature, $T$, using \eqref{eq:kn_smarr} and \eqref{eq:k_T},
    \begin{align}
        T=\frac{-4\pi^4 J^2 l^4-\pi^2 Q^2 l^2\left(\pi^2 Q^2 l^2+2S^2\right)+S^2\left(\pi l^2-(1+\epsilon)S\right)\left(\pi l^2-3(1+\epsilon)S\right)}{4\pi^{\frac{3}{2}}\sqrt{S^3(4\pi^3 J^2 l^2(\pi l^2-(1+\epsilon)S)+(\pi l^2(\pi Q^2+S)-(1+\epsilon)S^2)^2}}.
    \label{eq:kn_T}
    \end{align}
We then proceed to construct part of the relation by multiplying the inverse of \eqref{eq:kn_paep} and \eqref{eq:kn_T},
    \begin{align}
        -T\left(\frac{\partial S}{\partial \epsilon}\right)_{M,J,Q}=\frac{-4\pi^3 J^2 l^2 S^2-2\pi l^2 S^3(\pi Q^2+S)+2(1+\epsilon)S^5}{4\pi^{\frac{3}{2}}l^2\sqrt{S^3\left(4\pi^3 J^2 l^2(\pi l^2-(1+\epsilon)S)+(\pi l^2(\pi Q+S)-(1+\epsilon)S^2)^2\right)}}.
    \label{eq:kn_rh}
    \end{align}
Setting $T=0$ to obtain the Nariai limit condition, as in the previous subsection,
    \begin{align}
        -4\pi^4 J^2 l^4-\pi^2 Q^2 l^2\left(\pi^2 Q^2 l^2+2S^2\right)+S^2\left(\pi l^2-(1+\epsilon)S\right)\left(\pi l^2-3(1+\epsilon)S\right)=0.
    \label{eq:kn_cal}
    \end{align}
Solving \eqref{eq:kn_cal}, we obtain four roots
    \begin{multline*}
        S_{1,2}=\frac{\pi l^2}{3(1+\epsilon)}+\frac{1}{2}\sqrt{\frac{2\pi^4 l^4}{9(1+\epsilon)^2}+\frac{4\pi^2 Q^2 l^2}{9(1+\epsilon)^2}+X} \\ \pm\frac{1}{2}\sqrt{\frac{4\pi^2 l^4}{9(1+\epsilon)^2}-\frac{8\pi^2 Q^2 l^2}{9(1+\epsilon)^2}-X+\frac{\frac{16\pi^3 l^6}{27(1+\epsilon)^3}+\frac{32\pi^3 Q^2 l^4}{9(1+\epsilon)^3}}{4\sqrt{\frac{2\pi^4 l^4}{9(1+\epsilon)^2}+\frac{4\pi^2 Q^2 l^2}{9(1+\epsilon)^2}+X}}},
    \end{multline*}
    
    \begin{multline*}
        S_{3,4}=\frac{\pi l^2}{3(1+\epsilon)}-\frac{1}{2}\sqrt{\frac{2\pi^4 l^4}{9(1+\epsilon)^2}+\frac{4\pi^2 Q^2 l^2}{9(1+\epsilon)^2}+X} \\ \pm\frac{1}{2}\sqrt{\frac{4\pi^2 l^4}{9(1+\epsilon)^2}-\frac{8\pi^2 Q^2 l^2}{9(1+\epsilon)^2}-X-\frac{\frac{16\pi^3 l^6}{27(1+\epsilon)^3}+\frac{32\pi^3 Q^2 l^4}{9(1+\epsilon)^3}}{4\sqrt{\frac{2\pi^4 l^4}{9(1+\epsilon)^2}+\frac{4\pi^2 Q^2 l^2}{9(1+\epsilon)^2}+X}}},
    \end{multline*}
where
    \begin{align*}
        \begin{split} &
            X=\frac{(\pi^2 l^4-2\pi^2 Q^2 l^2)^2-36(1+\epsilon)^2(4\pi^4 J^2 l^4+\pi^4 Q^4 l^4)}{9(1+\epsilon)^2Y}+\frac{Y}{9(1+\epsilon)^2},\\ &
            Y=\pi^2 l^6\left(Z^3-108(7l^2+2Q^2)A+\sqrt{(-(Z^2-36S)^3+(Z^3-216l^2A+108ZA)^2)^{\frac{1}{3}}}\right),\\ &
            Z=l^2-2Q^2, \\ &
            A=(4J^2+Q^4)(1+\epsilon)^2.
        \end{split}
    \end{align*}
Like in the KdS case, three roots, $S_1, S_3, S_4$ cannot revert to the SdS case if we set $Q=0, J=0$ as the limits. Therefore, $S_2$ is the Nariai condition of the KNdS black hole,
    \begin{multline}
        S_N=S_2=\frac{\pi l^2}{3(1+\epsilon)}+\frac{1}{2}\sqrt{\frac{2\pi^4 l^4}{9(1+\epsilon)^2}+\frac{4\pi^2 Q^2 l^2}{9(1+\epsilon)^2}+X} \\ -\frac{1}{2}\sqrt{\frac{4\pi^2 l^4}{9(1+\epsilon)^2}-\frac{8\pi^2 Q^2 l^2}{9(1+\epsilon)^2}-X+\frac{\frac{16\pi^3 l^6}{27(1+\epsilon)^3}+\frac{32\pi^3 Q^2 l^4}{9(1+\epsilon)^3}}{4\sqrt{\frac{2\pi^4 l^4}{9(1+\epsilon)^2}+\frac{4\pi^2 Q^2 l^2}{9(1+\epsilon)^2}+X}}}.
    \label{eq:kn_ext}
    \end{multline}
Substituting \eqref{eq:kn_ext} into \eqref{eq:kn_rh},
    \begin{align}
        \lim_{M \to M_N}\left(-T\frac{\partial S}{\partial\epsilon}\right)_{M}=\frac{-4\pi^3 J^2 l^2 S_2^2-2\pi l^2 S^3(\pi Q^2+S_2)+2(1+\epsilon)S_2^5}{4\pi^{\frac{3}{2}}l^2\sqrt{S_2^3\left(4\pi^3 J^2 l^2(\pi l^2-(1+\epsilon)S_2)+(\pi l^2(\pi Q+S_2)-(1+\epsilon)S_2^2)^2\right)}},
    \label{eq:kn_rhcom}
    \end{align}
which corresponds to the right side of \eqref{GP}. Finally, by substituting \eqref{eq:kn_ext} into \eqref{eq:kn_m} and taking the partial derivative with respect to $\epsilon$,
    \begin{align}
        \left(\frac{\partial M}{\partial \epsilon}\right)_{T,Q,J}=\frac{-4\pi^3 J^2 l^2 S_2^2-2\pi l^2 S^3(\pi Q^2+S_2)+2(1+\epsilon)S_2^5}{4\pi^{\frac{3}{2}}l^2\sqrt{S_2^3\left(4\pi^3 J^2 l^2(\pi l^2-(1+\epsilon)S_2)+(\pi l^2(\pi Q+S_2)-(1+\epsilon)S_2^2)^2\right)}}.
    \label{eq:kn_lfcom}
    \end{align}
We can thus confirm the consistency of \eqref{eq:kn_rhcom} and \eqref{eq:kn_lfcom}. This reveals that the GP relation remains valid for the Nariai limit of KNdS black holes, demonstrating that the GP relation is reasonably universal regardless of the variables included.

\section{Summary}
Goon and Penco investigated the thermodynamic relation of perturbed extremal black holes and its leading-order expansion with respect to the perturbation parameter \cite{Goon:2019faz}. Their theoretical framework bears a resemblance to the proof of the WGC derived from higher-derivative corrections on the action. Consequently, they argue that their relation can elucidate the WGC. In this study, we replicate their calculations with dS black holes in the Nariai limit.

Initially, we investigated SdS black holes, applying the Nariai limit as SdS black holes possess $r_o$ and $r_c$. Our results, expressed in \eqref{eq:sch_rhcom} and \eqref{eq:sch_lf}, share the same analytic solution, enabling us to establish the same relation as \eqref{GP}. This reaffirms the validity of the concept presented in \cite{Goon:2019faz} for the SdS case. Moving on to RNdS black holes, with three horizons $r_i$, $r_o$ and $r_c$, we selected the extremality condition to align with the Nariai limit. Our calculations yielded identical results in \eqref{eq:rn_rhcom} and \eqref{eq:rn_lf}, verifying the applicability of \eqref{GP} to RNdS black holes with the Nariai limit. Finally, KdS black holes, featuring more than two solutions similar to the RNdS cases, were studied. In the case of KdS, we utilized \eqref{eq:Smarr} to obtain variables. With certain assumptions, we successfully reproduced the results discussed in the subsection explaining KdS. Through various processes, we established that \eqref{GP} could be derived in KdS black holes. Furthermore, considering that our computations are based on the Nariai limit of the dS spacetime, the presence of a negative sign in exact solutions such as \eqref{eq:sch_lf}, \eqref{eq:rn_lf}, and \eqref{eq:k_lf} can be attributed to the mass bound. Moreover, since the negative sign is a consequence of the positive cosmological constant, it is closely linked to the dS spacetime. As our results share the same form as \eqref{GP}, we can extend our analysis to \eqref{GPapp}. This implies a proportional relation between the shifts in mass and entropy, as indicated in \cite{Cheung:2018cwt, Cheung:2019cwi, Kats:2006xp}.

In summary, our calculations, at least for the studied black holes, reveal the universal nature of the relations. The exploration of the proportional relationship between corrected mass and entropy provides a pathway to understanding the WGC. The WGC offers intriguing insights into the realm of quantum gravity, and studies such as the one presented in this paper contribute to a deeper comprehension of this fundamental aspect of physics.

\vspace{10pt} 

\noindent{\bf Acknowledgments}

\noindent This research was supported by Basic Science Research Program through the National Research Foundation of Korea (NRF) funded by the Ministry of Education (NRF-2022R1I1A2063176) and the Dongguk University Research Fund of 2023. BG appreciates APCTP for its hospitality during completion of this work.\\

\bibliographystyle{jhep}
\bibliography{ref_v2}
\end{document}